\documentclass[11pt]{article}
\usepackage[T1]{fontenc}
\usepackage[utf8]{inputenc}
\usepackage{amsmath,amssymb}
\usepackage{newtxtext,newtxmath}
\usepackage[a4paper,margin=20mm]{geometry}
\usepackage{graphicx}
\usepackage{microtype}
\usepackage[font=small,labelfont=bf,singlelinecheck=false]{caption}
\DeclareCaptionLabelSeparator{bar}{\space|\space}
\usepackage[authoryear,round]{natbib}
\usepackage{xurl}
\usepackage[hidelinks,unicode]{hyperref}
\hypersetup{pdftitle={Anomalous pressure-dependent viscosity of basaltic melts and its role in asthenosphere melt accumulation},pdfauthor={Hongkun Zeng, Peiyu Zhang, Liang Yuan, Youjun Zhang, Xiang Wu, Junfeng Zhang}}
\DeclareUnicodeCharacter{03B1}{\ensuremath{\alpha}}
\DeclareUnicodeCharacter{03B2}{\ensuremath{\beta}}
\DeclareUnicodeCharacter{03B4}{\ensuremath{\delta}}
\DeclareUnicodeCharacter{03B7}{\ensuremath{\eta}}
\DeclareUnicodeCharacter{03C1}{\ensuremath{\rho}}
\DeclareUnicodeCharacter{03C3}{\ensuremath{\sigma}}
\DeclareUnicodeCharacter{03C4}{\ensuremath{\tau}}
\DeclareUnicodeCharacter{03C6}{\ensuremath{\varphi}}
\DeclareUnicodeCharacter{03BC}{\ensuremath{\mu}}
\DeclareUnicodeCharacter{0394}{\ensuremath{\Delta}}
\DeclareUnicodeCharacter{03F5}{\ensuremath{\epsilon}}
\DeclareUnicodeCharacter{221E}{\ensuremath{\infty}}
\DeclareUnicodeCharacter{2265}{\ensuremath{\geq}}
\DeclareUnicodeCharacter{2264}{\ensuremath{\leq}}
\DeclareUnicodeCharacter{2248}{\ensuremath{\approx}}
\DeclareUnicodeCharacter{2212}{\ensuremath{-}}
\DeclareUnicodeCharacter{2202}{\ensuremath{\partial}}
\DeclareUnicodeCharacter{2009}{\,}
\DeclareUnicodeCharacter{202F}{\,}
\DeclareUnicodeCharacter{2011}{-}
\DeclareUnicodeCharacter{2010}{-}
\DeclareUnicodeCharacter{27E8}{\ensuremath{\langle}}
\DeclareUnicodeCharacter{27E9}{\ensuremath{\rangle}}
\DeclareUnicodeCharacter{00D7}{\ensuremath{\times}}
\DeclareUnicodeCharacter{00B1}{\ensuremath{\pm}}

\providecommand{\doi}[1]{doi: {\urlstyle{same}\href{https://doi.org/#1}{\nolinkurl{#1}}}}

\begin{document}
\begin{center}
{\LARGE \textbf{Anomalous pressure-dependent viscosity of basaltic melts and its role in asthenosphere melt accumulation}\par}
\vspace{1em}
Hongkun Zeng\textsuperscript{1}, Peiyu Zhang\textsuperscript{1}, Liang Yuan\textsuperscript{1,3}, Youjun Zhang\textsuperscript{2}, Xiang Wu\textsuperscript{1}, Junfeng Zhang\textsuperscript{1}\par
\vspace{0.7em}
{\small \textsuperscript{1}State Key Laboratory of Geological Processes and Mineral Resources, School of Earth Sciences, China University of Geosciences, Wuhan, China; \textsuperscript{2}Institute of Atomic and Molecular Physics,~Sichuan University, Chengdu, China; \textsuperscript{3}Bayerisches Geoinstitut, Universität Bayreuth, Bayreuth, Germany.\par}
\vspace{0.6em}
Correspondence: L. Yuan (liang.yuan@uni-bayreuth.de)
\end{center}
\subsection*{Key points}
\begin{itemize}
\item 
Quantum-mechanical calculations reveal a pressure-induced viscosity minimum in basalt, dispelling recent concerns of experimental artifacts.

\item 
The viscosity minimum results from aluminum species relaxing stress efficiently, while the silicon--oxygen structural motifs stay rigid.

\item 
Melt mobility drops with ascent, trapping magma below the lithosphere and producing asthenosphere seismic anomalies.

\end{itemize}
\subsection*{Keywords}
Asthenosphere; seismic discontinuity; silicate melt; high pressure; viscosity

\subsection*{Abstract}
The asthenosphere's mechanical weakness enables plate tectonics, but its origin is debated. Partial melt has been proposed to cause this softening, yet recent studies suggest that the measured viscosity minimum in basaltic melts---an essential control on melt mobility---is an experimental artifact. Using quantum mechanics-based, machine learning-accelerated molecular dynamics, we extend simulation timescales by \textgreater10\textsuperscript{3} and achieve percent-level precision. We show that basaltic melt exhibits a robust viscosity minimum (\textasciitilde20\% below 1-bar values) at \textasciitilde3\,GPa, driven by pressure-induced reorganization of aluminum coordination that facilitates shear relaxation while silicon--oxygen polyhedra remain structurally rigid. Our results reveal a depth-dependent rheological transition: melt mobility peaks below \textasciitilde150\,km, promoting efficient extraction, but declines sharply during ascent, causing melt to stagnate beneath the lithosphere. This mechanism provides a physical basis for the dual seismic signatures of a melt-depleted deep asthenosphere and a melt-enriched layer near the lithosphere--asthenosphere boundary.

\subsection*{Plain language summary}
Earth's tectonic plates move because the underlying asthenosphere is uniquely soft. Scientists have long proposed that trace amounts of melt lubricate this layer, but recent studies suggested that earlier measurements of magma viscosity might be experimental artifacts. Using computer simulations based on quantum physics, we show that magma indeed becomes significantly less viscous at depths near 90 km. At these pressures, aluminum atoms reorganize into flexible coordination states that enhance melt mobility. However, as magma rises and temperature decreases, its viscosity increases dramatically. This viscosity increase impedes upward migration, causing melt to accumulate beneath the rigid lithosphere. Our findings explain the seismically detected melt-rich layer at the plate base and provide a mechanism for the rheological transition that defines the asthenosphere's lower boundary at \textasciitilde150 km depth.


\section{Introduction}\label{sec:1}

Silica-poor, low-viscosity basaltic lavas drive over 90\% of Earth's magmatism. In the upper mantle---at pressures (\emph{P}) \textgreater10\textsuperscript{4} bar---basaltic melt remains mobile yet undergoes dramatic structural and rheological transitions. Most strikingly, contrary to the conventional expectation that compression stiffens materials, its shear viscosity (\emph{η}) does not increase with \emph{P}. Instead, \emph{η} drops to a minimum at depths of \textasciitilde90--150\,km before rising again. This negative \emph{P--η} trend, observed in both experiments (\citealp{Sakamaki2013}) and simulations (\citealp{Dufils2017}), underpins key asthenospheric processes, including magma transport (\citealp{Sakamaki2013}), seismic anomalies (\citealp{Hua2023}; \citealp{Schmerr2012}), and geochemical stratification (\citealp{Zhang2024}).

Recently, however, the existence of this \emph{η} minimum---and the associated negative \emph{P--η} regime---has been challenged. \citet{Russell2025} contend that the signal reported by \citet{Sakamaki2013} is an experimental artifact, arising from the absence of 1-atm reference data and unaccounted temperature (\emph{T}) fluctuations of ±80\,K; once corrected, they argue, \emph{η} increases monotonically with \emph{P} (at constant \emph{T}).

This reinterpretation exposes a central and often underappreciated challenge in high \emph{P--T} viscometry: \emph{P} and \emph{T} effects cannot be cleanly disentangled. For instance, Paris--Edinburgh press \emph{η} measurements (\citealp{Kono2014}) rely on power--temperature calibrations rather than direct thermometry, introducing ≥50 K of uncertainty---sufficient to obscure the subtle \emph{P--η} curvature (\citealp{Russell2025}). Comparable \emph{T} uncertainties also afflict multi-anvil experiments, even those using thermocouples (e.g., COMPRES assemblies; \citealp{Leinenweber2012}). More critically, the falling sphere method---the principal \emph{in situ} technique for high \emph{P--T} viscosity measurements---suffers from wall-induced drag in millimeter-scale sample capsules, which violates Stokes' law; to date, no universally accepted correction for this effect exists (\citealp{Ashley2024}).

Given these experimental limitations, computational approaches may offer the only viable path to resolve whether basaltic melts truly exhibit an \emph{η} minimum in the upper mantle. Here, we use density functional theory (DFT)-based molecular dynamics (MD), accelerated by machine learning, to access microsecond timescales---three orders of magnitude beyond standard DFT-MD. This capability allows us to resolve the \emph{P} dependence of basaltic melt \emph{η} with high precision, reducing statistical uncertainties to \textasciitilde1--3\% (from up to 50\% in prior studies), and to cleanly disentangle \emph{P} and \emph{T} effects under tightly controlled conditions.

\section{Methods}\label{sec:2}

Over the past decades, \emph{ab initio} liquid \emph{η} has been largely computed using DFT (e.g., \citealp{Alfe1998}), but its cubic scaling limits simulations to small systems and short times, hindering convergence of transport properties. Neural network potentials (NNPs; \citealp{Behler2007}) overcome this by reproducing DFT energies, forces, and stresses at far lower cost. Here, NNP-based MD enables \emph{η} calculations with substantially reduced statistical uncertainty.

We constructed NNPs for the Di\textsubscript{64}An\textsubscript{36} eutectic composition (CaMgSi\textsubscript{2}O\textsubscript{6}--CaAl\textsubscript{2}Si\textsubscript{2}O\textsubscript{8}), a widely used model basalt (23.5 wt\% CaO, 10.7 wt\% MgO, 15.5 wt\% Al\textsubscript{2}O\textsubscript{3}, 50.3 wt\% SiO\textsubscript{2}; e.g., \citealp{Asimow2010}), using DeePMD-kit (\citealp{Wang2018}). Following our earlier work (\citealp{Yuan2023}; \citealp{ZhangP2025}), we used DP-GEN active learning (\citealp{Zhang2020}) to expand the training set across 0--40 GPa and 0--4000 K. Representative CaO--MgO--Al\textsubscript{2}O\textsubscript{3}--SiO\textsubscript{2} (CMAS) minerals were additionally included to improve configurational diversity and validate crystalline densities.

NNP-MD simulations were performed with LAMMPS (\citealp{Thompson2022}). Melts were first equilibrated under target \emph{P}--\emph{T} conditions in the isothermal--isobaric (\emph{NPT}) ensemble, followed by isothermal--isochoric (\emph{NVT}) production runs for \emph{η} calculations. A Nosé--Hoover barostat/thermostat (\citealp{Hoover1985}; \citealp{Nose1984}) and a 2.0\,fs time step were used for integration.

\section{Results}\label{sec:3}

\subsection{NNP validation}\label{sec:3.1}

To validate the NNP beyond its excellent agreement with DFT test data, we compared calculated mass densities (\emph{ρ}\textsubscript{cal}) of representative minerals in the CMAS system against experimental values (\emph{ρ}\textsubscript{exp}) (Figure~\ref{fig:1}a). Crystal \emph{ρ} from x-ray diffraction provides a stringent benchmark, as liquid \emph{ρ} measurements under melting conditions remain highly uncertain. The SCAN-NNP reproduces experimental crystal \emph{ρ} with exceptional accuracy, yielding relative errors (\emph{ρ}\textsubscript{cal} - \emph{ρ}\textsubscript{exp})/\emph{ρ}\textsubscript{exp} of Fo, +0.23\%; En, +0.66\%; Pe, -0.53\%; Sp, -0.31\%; Py, -0.40\%; Gr, +0.04\%; Dvm, -0.95\%; Cor, -0.12\%; An, -0.69\%; and Di, +0.25\%. By comparison, the LDA-NNP overestimates \emph{ρ} (+0.96\% to +3.02\%), while the GGA-NNP underestimates them (-3.24\% to -4.69\%). These trends reflect known exchange--correlation biases, where LDA typically underestimates and GGA overestimates lattice parameters (\citealp{Perdew2008}).

For high-\emph{T} melts, we compared the equation of state (EOS) of liquid Di\textsubscript{64}An\textsubscript{36} from NNP-MD with thermodynamic predictions from MAGEMin (\citealp{Riel2022}), using the thermodynamic databases of \citet{Holland2018} and \citet{Green2025}. At 0 GPa and 1673 K, MAGEMin predicts a melt \emph{ρ} of 2.62 g/cm\textsuperscript{3}, consistent with oxide partial molar volumes (\citealp{Lange1997}). SCAN-NNP matches this within -2.04\% (equivalent to a +0.45 GPa \emph{P} overestimate), whereas LDA- and GGA-NNPs deviate by +4.99\% (-0.78 GPa) and -10.58\% (+2.25 GPa), respectively (Figure~\ref{fig:1}b). The EOS of \citet{Martin2012} agrees with ours despite using an empirical potential, and the shock-compression EOS of \citet{Asimow2010}, anchored to the same \citet{Lange1997} reference \emph{ρ}, deviates only slightly at higher \emph{P}. The \citet{Vuilleumier2009} \emph{ρ} at 2273 K, adjusted to 1673 K via the thermal expansion coefficient \emph{α} = 6.084 × 10\textsuperscript{-5} K\textsuperscript{-1} (\citealp{Lange1997}), also falls close to the MAGEMin reference. To correct the systematic bias of SCAN-NNP, we fitted the \emph{P} difference \(\Delta P_{\mathrm{SCAN-Ref}}\mathrm{(GPa)} = P_{\mathrm{SCAN}}-P_{\mathrm{Ref}}\) (where \emph{P}\textsubscript{SCAN} and \emph{P}\textsubscript{Ref} are pressures from SCAN-NNP simulations and MAGEMin, respectively, evaluated at identical densities) with a stretched exponential \(0.45\exp\left[-0.24P_{\mathrm{SCAN}}^{2.53}\right]\), preserving monotonicity and non-negative extrapolation. We apply the 1673 K \emph{P} correction at all simulation \emph{T}, assuming a \emph{T}-independent offset at fixed \emph{ρ}. Offsets are broadly consistent across 1673--1873 K; 2073 K is extrapolated.

\subsection{Liquidus curve}\label{sec:3.2}

Following experimental constraints (Table~1 of \citealp{Wang2014}), we focused on the 0--6 GPa \emph{P}-range, where a negative \emph{P--η} trend has been repeatedly reported. The exact \emph{P} at which \emph{η} reaches a minimum is secondary; instead, the key requirement is to accurately capture the initial negative slope before \emph{η} increases again at higher \emph{P}. Because this anomaly strengthens at lower \emph{T} (\citealp{Dufils2017}), we first determined the liquidus temperature (\emph{T}\textsubscript{L}) of the Di\textsubscript{64}An\textsubscript{36} basalt across this \emph{P} range.

At 1 atm, Di\textsubscript{64}An\textsubscript{36} lies near the Di--An eutectic (\textasciitilde1573 K), where phase relations are well constrained (\citealp{Bowen1915}; \citealp{Osborn1942}). Its high-\emph{P} melting behavior was evaluated using CMAS phase-equilibrium modeling with MAGEMin (\citealp{Riel2022}), which reproduces the 1-atm liquidus and predicts a rise to \textasciitilde2073 K at 5 GPa. We therefore calculated \emph{η} along two paths: (1) isothermal 2073 K to isolate \emph{P} effects in a fully molten state, and (2) a near-liquidus path slightly above the equilibrium liquidus to mimic experimental conditions where \emph{T} necessarily increases with \emph{P}.

Structure and dynamics analyses confirm that Di\textsubscript{64}An\textsubscript{36} remains liquid along both paths. Linear mean-squared displacements (MSD) growth indicates diffusive behavior, with Si showing the lowest mobility among cations, consistent with DFT-MD results (\citealp{deKoker2008}). The radial distribution functions, \emph{g}(\emph{r}), show liquid-like short-range order with damped oscillations approaching unity, unlike crystalline diopside and anorthite, which exhibit sharp periodic peaks from long-range order.

\subsection{Shear viscosity}\label{sec:3.3}

\emph{Viscosity from the Green}--\emph{Kubo relation}. Shear viscosity was computed via the Green--Kubo relation (\citealp{Allen2017}) using equilibrium MD simulations driven by the SCAN-trained NNP. The shear viscosity \emph{η} is given by:

\begin{equation}\label{eq:1}
\eta = \frac{V}{k_{\mathrm B}T}\int_0^\infty\left\langle\sigma_{\alpha\beta}(t_0+\Delta t)\sigma_{\alpha\beta}(t_0)\right\rangle\,\mathrm{d}t,
\end{equation}

where \emph{V} is the system volume, \emph{k}\textsubscript{B} is the Boltzmann constant, \emph{σ\textsubscript{αβ}} denotes the off-diagonal stress tensor components (\emph{P\textsubscript{xy}}, \emph{P\textsubscript{xz}}, \emph{P\textsubscript{yz}}). The angular brackets represent an average over all time origins \emph{t}\textsubscript{0}. To improve statistical convergence, we also included shear contributions from diagonal stress differences, (\emph{P\textsubscript{xx}} -- \emph{P\textsubscript{yy}})/2 and (\emph{P\textsubscript{yy}} -- \emph{P\textsubscript{zz}})/2, following \citet{Alfe1998}.

The stress autocorrelation function (ACF) at 1619 K and 0 GPa decays to zero rapidly (within 300 ps). In principle, its time integral should converge to a well-defined plateau corresponding to \emph{η}. In practice, however, statistical fluctuations---arising from noise in the ACF that resembles discrete white noise---accumulate upon integration, producing a random walk that obscures any clear plateau (\citealp{Oliveira2017}).

To quantify this challenge, we computed \emph{η} from ensembles of 150--600 independent \emph{NVT} trajectories, each 5 ns long. All simulations at a given \emph{P--T} condition started from the same equilibrated configuration but with distinct random initial velocities. We identify two salient features: (1) individual \emph{η}(\emph{t}) curves lack a definitive plateau; instead, they begin to diverge after 100 ps and exhibit a spread exceeding 100\% relative variation by 2000 ps (i.e., 40\% of the trajectory length); and (2) while ensemble averaging suppresses noise and reveals an overall trend, residual fluctuations persist, and the mean \emph{η}(\emph{t}) remains marginally unstable even at the longest simulation times.

These substantial statistical uncertainties can mask subtle \emph{P}-dependent effects. For instance, the \textasciitilde50\% reduction in \emph{η} reported by \citet{Sakamaki2013}---from 0.286 Pa s at 3.14 GPa to 0.151 Pa s at 4.49 GPa (both at \textasciitilde2000 K)---falls entirely within the noise envelope of individual Green--Kubo estimates. Resolving the proposed low-\emph{P} \emph{η} minimum therefore requires rigorous error control.

To obtain statistically robust \emph{η} values, we used the time decomposition method (\citealp{Zhang2015}). We (1) generated \emph{M} independent 5-ns \emph{NVT} trajectories, (2) computed \emph{η}(\emph{t}) for each trajectory via the Green--Kubo relation (Eq.~\ref{eq:1}; Figure~\ref{fig:1}c), (3) calculated the ensemble-averaged viscosity \(\left\langle\eta(t)\right\rangle\) and its time-dependent standard deviation \emph{σ}(\emph{t}) (Figure~\ref{fig:1}d), (4) fitted \emph{σ}(\emph{t}) to a power law, \emph{σ}(\emph{t}) = \emph{A}\textsubscript{0}\emph{t\textsuperscript{b}}, and (5) fitted \(\left\langle\eta(t)\right\rangle\) to a double-exponential function over the interval {[}0, \emph{t}\textsubscript{cut}{]} (Figure~\ref{fig:1}e). The cutoff time \emph{t}\textsubscript{cut} is defined where \emph{σ}(\emph{t}) reaches 20--50\% of the approximate plateau value of \(\left\langle\eta(t)\right\rangle\); sensitivity tests using this range of \emph{t}\textsubscript{cut} values produce statistically indistinguishable final \emph{η} values. Before fitting, we define the plateau estimate as the ensemble-averaged running Green--Kubo integral at a 2 ns correlation lag, the reference for the \emph{σ}(\emph{t}) cutoff criterion. A weighting factor of 1/\emph{t\textsuperscript{b}} was applied during fitting to give greater influence to reliable short-time data while downweighting noisy long-time data. The long-time limit of the fitted \(\left\langle\eta(t)\right\rangle\) yields the final viscosity, \emph{η}\textsuperscript{∞}, with uncertainties estimated via bootstrap standard errors.

By performing large ensembles of replicas (\emph{M} ≥ 150), we constrain statistical uncertainties in \emph{η}\textsuperscript{∞} to 1--3\%, and system size effects are small: \emph{η}\textsuperscript{∞} from the baseline system (\emph{N} = 976) is 3.81\% and 3.72\% larger than those of larger systems (\emph{N = 1952 and 3904)}, consistent with the negligible system size dependence reported for Green--Kubo \emph{η} calculations (\citealp{Alfe1998}). Both sources of uncertainty are minor compared with the \emph{P}-induced \emph{η} change (\textasciitilde23\%). Adequate configurational sampling is supported by the MSD of the slowest species (Si). Even under high-\emph{P} conditions that suppress mobility (6 GPa, 2073 K), combining 150 independent trajectories yields \(\sqrt{\mathrm{MSD}_{\mathrm{Si}}}/\sqrt[3]{V}\) \textgreater{} 40, nearly two orders of magnitude larger than values reported in comparable DFT-MD studies (\textasciitilde0.4; \citealp{Bajgain2022}; \citealp{Majumdar2020}), suggesting thorough configurational sampling.

Figures~\ref{fig:2}a,b compare \emph{η}\textsuperscript{∞} along two \emph{P--T} paths: an isotherm at 2073 K and the liquidus. Along the isotherm, \emph{η} exhibits a U-shaped dependence---decreasing from 0.1080 ± 0.0010 Pa s at 0 GPa to a minimum of 0.0879 ± 0.0011 Pa s at 3 GPa, then increasing to 0.1035 ± 0.0013 Pa s at 6 GPa. This indicates a negative compressional response below 3 GPa, followed by the expected \emph{η} increase at higher \emph{P}. By contrast, \emph{η} along the liquidus decreases by a factor of \textasciitilde23---from 1.9310 ± 0.0383 Pa s at 0 GPa to 0.0835 ± 0.0009 Pa s at 6 GPa. This steep negative \emph{P--η} trend reflects thermal effects rather than intrinsic \emph{P} dependence: at low \emph{P}, the isotherm samples a far more superheated melt (higher \emph{T}/\emph{T}\textsubscript{L}) than the liquidus. The contrast is most pronounced at 0 GPa, where \emph{η} increases nearly eighteenfold---from 0.1080 ± 0.0010 Pa s at 2073 K to 1.9310 ± 0.0383 Pa s at 1619 K, which is slightly above the liquidus (Δ\emph{η}/Δ\emph{T} ≈ -0.004 Pa s K\textsuperscript{-1}). Typical experimental \emph{T} uncertainties (\textasciitilde100 K) would thus alter \emph{η} by \textasciitilde0.4 Pa s---over twenty times the amplitude of the U-shaped isothermal signal---explaining why this subtle feature has eluded detection in previous experiments. These contrasting isothermal and liquidus behaviors are consistent with recent \emph{in situ} measurements (\citealp{Xie2020}), which show that the negative \emph{P--η} slope observed along the liquidus weakens---and can even reverse---under isothermal conditions.

\emph{Parameterization and physical insights.} We describe the computed \emph{η} results along both \emph{P--T} paths using an Arrhenius relation (e.g., \citealp{Hui2007}),

\begin{equation}\label{eq:2}
\eta(P,T) = \eta_0\exp\left(\frac{E_a}{RT}\right),
\end{equation}

in which the activation energy \emph{E}\textsubscript{a} varies with \emph{P}, and \emph{R} is the gas constant. The pre-exponential factor \emph{η}\textsubscript{0} is a global constant, independent of \emph{P} and \emph{T}, so \emph{P} dependence enters solely through \emph{E}\textsubscript{a}(\emph{P}). Representing \emph{E}\textsubscript{a}(\emph{P}) as a quadratic function, \(E_a(P) = E_0+E_1P+E_2P^2\), yields an excellent fit to the data (Figure~\ref{fig:2}c; \emph{R}\textsuperscript{2} = 0.99). Although this parameterization describes the results over 0--6 GPa and 1619--2223 K, it does not explain the mechanism underlying the negative \emph{P--η} trend.

To elucidate the physical basis of this \emph{P}-induced \emph{η} reduction, we invoke the Maxwell model, \(\eta = G^\infty\tau_{\mathrm M}\), which decomposes \emph{η} into the instantaneous shear modulus \emph{G}\textsuperscript{∞} and the relaxation time \emph{τ}\textsubscript{M}. In experiments, \emph{G}\textsuperscript{∞} corresponds to the shear modulus measured at high acoustic frequencies (typically MHz--GHz). In MD simulations, this quantity maps onto the plateau shear modulus, \emph{G}\textsubscript{P}, that appears at intermediate times in the shear stress ACF, \(G(\Delta t) = \frac{V}{k_{\mathrm B}T}\langle\sigma_{\alpha\beta}(t_0+\Delta t)\sigma_{\alpha\beta}(t_0)\rangle\) (\citealp{Puosi2012}). We estimated \emph{G}\textsubscript{P} by fitting the 0.1--100 ps segment of \emph{G}(Δ\emph{t}) with a stretched exponential \(G_{\mathrm P}\exp\left[-(\Delta t/\tau_0)^\beta\right]\), where \emph{G}\textsubscript{P}, \emph{τ}\textsubscript{0} and \emph{β} are fitting parameters (\citealp{Flenner2019}).

This analysis reveals that \emph{G}\textsubscript{P} (\emph{G}\textsuperscript{∞}) stiffens systematically under compression, increasing from 22 GPa at \emph{P} = 0 GPa to 29 GPa at \emph{P} = 6 GPa, within the experimental range of 5--42 GPa (\citealp{Dingwell1989}). As \emph{G}\textsuperscript{∞} increases steadily with \emph{P}, the observed reduction in \emph{η} demands a correspondingly rapid fall in \emph{τ}\textsubscript{M}. Hence, the negative \emph{P--η} response arises not from elastic softening, but from accelerated structural relaxation, a mechanism we examine in greater detail in Section~\ref{sec:4.2}.

\section{Discussion}\label{sec:4}

\subsection{Previous related work}\label{sec:4.1}

The \emph{η} of upper-mantle partial melts has long been a focus of high \emph{P--T} experimentation, yet published datasets remain inconsistent: some report a clear isothermal negative \emph{P--η} relation (\citealp{Bonechi2022}; \citealp{Sakamaki2013}; \citealp{Zhou2024}), while others find no such effect (\citealp{Xie2020}). Interpreting \emph{η} variations with \emph{P} requires distinguishing between along-liquidus and isothermal conditions. The large \emph{η} decrease reported by \citet{Bonechi2022} reflects the combined effects of \emph{P} and \emph{T} along the liquidus, rather than \emph{P} alone---a trend also observed in our simulations (Figure~\ref{fig:2}b). For the isothermal results of \citet{Zhou2024}, \emph{η} decreases by at most \textasciitilde2.7× across the measured \emph{P} range for a given H\textsubscript{2}O content, comparable to the relatively narrow variation reported by \citet{Sakamaki2013}. Our simulated \emph{η} at 2073 K are systematically lower than those of \citet{Sakamaki2013} by \textasciitilde1 order of magnitude, likely reflecting both their lower measurement \emph{T} (\textasciitilde1850--2000 K) and compositional differences between our FeO- and Na\textsubscript{2}O-free, Ca-rich Di\textsubscript{64}An\textsubscript{36} melt and their basaltic composition (9.9 wt\% FeO, 2.5 wt\% Na\textsubscript{2}O, and lower CaO). Our results show a weak U-shaped \emph{P} dependence, contrasting with the sharper trend reported by \citet{Sakamaki2013} and the monotonic increase predicted by the isothermal model of \citet{Russell2025}. However, such comparisons should be interpreted cautiously, because falling sphere measurements are typically performed within a similar super-liquidus range (\textasciitilde100--200 K above the liquidus) rather than at strictly identical \emph{T} across \emph{P}, and they involve greater thermal-control challenges than along-liquidus measurements (\citealp{Cochain2017}; \citealp{Liebske2005}).

Although compositional differences among melts and \emph{T} variations contribute to the observed scatter, a primary source of discrepancy lies in the difficulty of accurately measuring \emph{η} under extreme conditions (\citealp{Ashley2024}). Theoretical calculations offer atomistic insights that may help reconcile these inconsistencies (\citealp{Wang2014}; \citealp{ZhangS2025}). However, computational predictions of \emph{η} are themselves subject to uncertainties. To address the dispute between \citet{Russell2025} and \citet{Sakamaki2013}, we focus on two major sources of error that have hindered previous theoretical studies: DFT pressure inaccuracies and statistical limitations in Green--Kubo integration. Both issues are well documented in materials and physical sciences (\citealp{Oliveira2017}; \citealp{Perdew2008}); here, we examine each in the context of modeling Earth's silicates.

\emph{Pressure} \emph{bias.} Before computing \emph{η} using DFT, one must address a fundamental source of uncertainty: pressure. For iron-free crystalline silicates, pressures predicted by LDA and GGA functionals can deviate from experimental values by several gigapascals (\citealp{Perdew2008}; Figure~\ref{fig:1}a). While such deviations are negligible in studies spanning the full mantle (0--130 GPa), they become consequential when resolving subtle \emph{η} variations at low \emph{P} (\textless5 GPa).

For high-\emph{T} melts, however, a systematic analysis of DFT pressure accuracy remains scarce, owing to limited experimental EOS data and narrow compositional coverage. Recent DFT-MD studies of basalt \emph{η} (\citealp{Bajgain2022}; \citealp{Majumdar2020}) relied on GGA functionals: \citet{Majumdar2020} did not report \emph{P} errors, and \citet{Bajgain2022} noted that their GGA-based EOS underestimated \emph{ρ} by \textasciitilde10\%, implying \emph{P} overestimates of several gigapascals. Yet compositional and \emph{P--T} differences between their simulations and experiments prevented a definitive quantification of these errors. To overcome the scarcity of high-\emph{T} melt \emph{ρ} data and enable a transparent evaluation of \emph{P} uncertainties, we constructed a thermodynamically consistent reference EOS and used it to calibrate raw DFT pressures at strictly matched \emph{P--T} conditions and melt composition. A key element of this strategy is the use of the SCAN meta-GGA functional (\citealp{Sun2015}), which---despite its greater computational expense---reduces \emph{P} errors to \textless0.5 GPa (Section~\ref{sec:3.1}).

\emph{Time convergence.} DFT-MD simulations of silicate melts typically span tens of picoseconds for static properties (e.g., \citealp{Stixrude2005}). Transport properties such as \emph{η}, however, require substantially longer trajectories. Computational constraints have generally limited such simulations to less than 100 ps, with only a few studies reaching 120--240 ps (\citealp{Majumdar2020}) or, at considerably greater computational expense, 400 ps (\citealp{Bajgain2022}). Despite this increase in trajectory length, the \emph{η} values and their \emph{P} dependence reported by Bajgain et al. differ markedly from those of Majumdar et al. for the same basalt composition---highlighting the challenge of achieving time-converged \emph{η} estimates with DFT-MD.

While previous \emph{ab initio} MD simulations confirmed that the stress ACF decays to zero, this alone does not guarantee \emph{η} convergence, as statistical noise can obscure the plateau in the time-integrated stress ACF (\citealp{Oliveira2017}). Following best practices for transport property calculations (\citealp{Maginn2019}), our results demonstrate that reliable \emph{η} estimates for basaltic melts require substantially longer, statistically independent sampling (Figures~\ref{fig:1}c--e). To achieve this, we ran extensive simulations on a 976-atom system over finely spaced \emph{P} from 0 to 6 GPa, generating 150--600 independent 5-ns trajectories per \emph{P--T} point---totaling ≥7.5 μs of sampling along each \emph{P--T} path. This massive computational investment---orders of magnitude beyond typical DFT-MD---ensures statistically reliable results that resolve the anomalous \emph{P} dependence of isothermal \emph{η}.

\subsection{Microscopic origins of the \emph{η} anomaly}\label{sec:4.2}

\emph{P}-driven anomalies in the transport properties of silicate melts are often attributed to changes in the coordination of network-forming cations (e.g., Si and Al). Previous computational studies reported a diffusivity maximum in NaAlSi\textsubscript{2}O\textsubscript{6} melt near 25 GPa, coinciding with a pronounced increase in fivefold Si--O coordination (\citealp{Angell1982}). Similarly, transient five-coordinated Si has been proposed to enhance viscous flow in MgSiO\textsubscript{3} melts (\citealp{Karki2010}).

By contrast, network modifier bonds (e.g., Ca, Mg) break and reform on ultrafast timescales (\textless50 fs for Mg--O; \textless30 fs for Ca--O; \citealp{Solomatova2019})---orders of magnitude faster than \emph{τ}\textsubscript{M} (\textasciitilde10\textsuperscript{3} fs; Section~\ref{sec:3.3}). Consequently, these modifier bonds cannot meaningfully influence viscous relaxation. We therefore focus our analysis exclusively on network formers (Si and Al), characterizing their local environments via coordination numbers (CNs): the number of oxygen neighbors within the first coordination shell, defined by a cutoff radius \emph{r}\textsubscript{cut} set at the first minimum of the Si(Al)--O \emph{g}(\emph{r}).

\emph{Limits of static Si(Al)--O coordination.} In contrast to the non-monotonic \emph{P} dependence of viscosity---\emph{η} decreases below 3 GPa and then increases---the CNs of Si and Al vary strictly monotonically with \emph{P} at 2073 K. Silicon remains overwhelmingly fourfold coordinated; the fraction of fivefold Si rises only modestly, from 0.83 at\% at 0 GPa to 6.33 at\% at 6 GPa. This subdued evolution is consistent with prior DFT-MD studies (e.g., \citealp{Stixrude2005}), which show that fivefold Si becomes dominant only at much higher \emph{P} (25--50 GPa)---far beyond the 0--6 GPa range where experimental \emph{η} minima occur (\citealp{Wang2014}). Aluminum exhibits a similarly monotonic shift: its 4-, 5-, and 6-fold populations evolve from 78.47 at\%, 19.79 at\%, and 1.22 at\% at 0 GPa to 33.40 at\%, 49.99 at\%, and 16.45 at\% at 6 GPa, respectively.

The steady rise in fivefold Si and Al stands in clear contrast to the U-shaped \emph{P--η} trend, showing that static CN populations alone cannot account for low-\emph{P} transport anomalies. This mismatch highlights a central limitation of static coordination statistics: they ignore dynamics. In silicate melts, Si(Al)--O bonds are not permanent structural units but transient linkages that continuously form and break, endowing each polyhedral configuration with a finite lifetime, \emph{τ}\textsubscript{CN}, which static analyses overlook.

\emph{Dynamics of Si(Al)--O exchange and stability.} To quantify the dynamics of Si(Al)--O environments, we computed \emph{τ}\textsubscript{CN} from the coordination ACF, following the intermittent formalism originally developed for hydrogen-bond lifetimes in water (\citealp{Luzar1996}):

\begin{equation}\label{eq:3}
\tau_{\mathrm{CN}} = \int_0^\infty\left\langle\frac{\sum_i h_i(t_0)\,h_i(t_0+\Delta t)}{\sum_i h_i(t_0)^2}\right\rangle\,\mathrm{d}t.
\end{equation}

Here, \emph{h\textsubscript{i}} = 1 only when the \emph{i}-th Si(Al) retains both its coordination number (4, 5, or 6) and the identical O-neighbor set at times \emph{t}\textsubscript{0} and \emph{t}\textsubscript{0}+Δ\emph{t}; otherwise \emph{h\textsubscript{i}} = 0. This intermittent definition allows temporary exits and re-entries of a coordination state, rather than irreversibly terminating bonds as in continuous definitions. To further suppress noise-induced bond jitter---which would otherwise fragment a single bond into many break/reform events (\citealp{Schulze2023})---we introduced a hysteresis buffer \(\delta=0.8\sigma_{g(r)}\) around the bond cutoff \emph{r}\textsubscript{cut}, where \(\sigma_{g(r)}\) is the Gaussian width of the first peak in the Si(Al)--O \emph{g}(\emph{r}). Bond states were updated according to three rules: (1) formation when \(r<r_{\mathrm{cut}}-\delta\), (2) breaking when \(r>r_{\mathrm{cut}}+\delta\), and (3) persistence (no change) in the intermediate regime \(|r-r_{\mathrm{cut}}|\leq\delta\).

Figures~\ref{fig:2}d,e compare the bulk relaxation times estimated from the Maxwell model, \emph{τ}\textsubscript{M}, with the coordination-specific \emph{τ}\textsubscript{CN} across 0--6 GPa. Fourfold Si--O units remain long-lived but shorten markedly---from 87 ps at 0 GPa to 33 ps at 6 GPa---while fivefold units, nearly absent at low \emph{P}, stabilize modestly to \textasciitilde1 ps at 6 GPa. Neither approaches \emph{τ}\textsubscript{M} (which decreases from 9 ps to 3 ps): SiO\textsubscript{4} lifetimes are orders of magnitude too long, whereas SiO\textsubscript{5} lifetimes are far too short. For Al, \emph{τ}\textsubscript{CN} decreases from 11 to 4 ps for AlO\textsubscript{4} and increases from 1 to 2 ps for AlO\textsubscript{5}, bracketing \emph{τ}\textsubscript{M}. Although AlO\textsubscript{6} grows substantially (from 1.22 at\% to 16.45 at\%), its lifetime remains brief (0--2 ps)---inconsistent with \emph{τ}\textsubscript{M}.

To assess average cation dynamics, we computed two ensemble-averaged lifetimes across 4-, 5-, and 6-fold species: a fraction-weighted mean, \(\tau_{\mathrm F} = \sum(f^i\tau_{\mathrm{CN}}^i)/\sum f^i\), and a rate-weighted (harmonic) mean, \(\tau_{\mathrm R} = 1/\sum(f^i/\tau_{\mathrm{CN}}^i)\), where \(f^i\) and \(\tau_{\mathrm{CN}}^i\) are the population fraction and lifetime of the \emph{i}-fold polyhedron (\emph{i} = 4, 5, 6). \emph{τ}\textsubscript{F}, dominated by long-lived fourfold Si and Al units, remains large (Si: 86--31 ps; Al: 9--3 ps). By contrast, \emph{τ}\textsubscript{R}---sensitive to fast-exchanging species---is substantially shorter (Si: 28--12 ps; Al: 4--3 ps). Both metrics decrease with \emph{P}, but only the Al-based \emph{τ}\textsubscript{R} matches \emph{τ}\textsubscript{M}. This correspondence indicates that macroscopic relaxation in the basalt melt is governed not by the dominant, slowly rearranging Si environments, but by the rarer, rapidly reconfiguring Al species---consistent with experimental evidence showing Al's disproportionate influence on silicate melt \emph{η} (\citealp{Giuliani2025}).

The \emph{τ}\textsubscript{R}(Al) correspondence with \emph{τ}\textsubscript{M} holds here only for Di\textsubscript{64}An\textsubscript{36}. Modifier field strength sets how strongly Al coordination responds to pressure in natural basaltic melts (\citealp{Kelsey2009}), and redox-sensitive Fe lowers \emph{η} upon reduction (\citealp{Chevrel2013}). FeO and Fe\textsubscript{2}O\textsubscript{3} act oppositely on Al--O coordination (\citealp{Ma2022}). Exchange rates depend on neither static populations nor bulk \emph{η}, and \emph{P} dependence is composition-dependent: polymerized melts turn over at 3--5 GPa, whereas depolymerized ones rise monotonically (\citealp{Wang2014}). The \textasciitilde3 GPa minimum is therefore not universal.

\subsection{Implications for asthenosphere melt migration}\label{sec:4.3}

Seismic discontinuities at depths of 80--100 km---corresponding to the Gutenberg discontinuity at the lithosphere--asthenosphere boundary (LAB; \citealp{Schmerr2012})---and near 150 km (\citealp{Hua2023}), together with constraints from geochemical Y/Yb ratios (\citealp{Zhang2024}), point to a vertically heterogeneous distribution of melt within the asthenosphere. This stratification has been interpreted as reflecting \emph{η} contrasts associated with mid-asthenospheric partial melts (\citealp{Sakamaki2013}). However, the reliability of these \emph{η} estimates has recently been questioned (\citealp{Russell2025}), casting doubt on both the existence of an \emph{η} anomaly and the inferred melt layering.

To investigate melt distribution in the asthenosphere, we constructed a simple melt transport model based on our new \emph{P--T--η} relation (Eq.~\ref{eq:2}). Melt migration through a permeable mantle matrix is governed by the two-phase extension of Darcy's law (\citealp{McKenzie1984}), which describes the relative motion between melt and solid and defines the melt segregation velocity:

\begin{equation}\label{eq:4}
V_{\mathrm D} = \frac{\Delta\rho\,g\,k_\varphi}{\eta\,\varphi},
\end{equation}

where Δ\emph{ρ} is the melt--solid density contrast, \emph{φ} is the porosity (melt fraction, assumed uniform at 10\textsuperscript{-3} across all depths), \emph{g} ≈ 9.8 m/s\textsuperscript{2} is gravitational acceleration, and \emph{k\textsubscript{φ}} is permeability.

Basaltic melts at depth contain dissolved volatiles, among which H\textsubscript{2}O exerts the strongest control on melt properties (\citealp{Giordano2008}). To isolate its effect, we calculate depth-dependent Δ\emph{ρ} for both wet and dry mantle conditions. Unlike earlier work that assumed a single-phase olivine (Mg\# = 90) and dry basalt (\citealp{Sakamaki2013}), we adopt a more geologically realistic melting model. The solid mantle is represented by KLB-1 peridotite (\citealp{Takahashi1986}) containing either 0 or 200 wt ppm H\textsubscript{2}O, corresponding to 0 or 3.33 wt\% H\textsubscript{2}O in the coexisting melt, based on a melt--peridotite H\textsubscript{2}O partition coefficient of 0.006 (\citealp{Tenner2009}). The melt phase is approximated by global mean basalt (\citealp{Gale2013}). Mineral and melt densities are computed using MAGEMin (\citealp{Riel2022}) along a 70-Myr oceanic plate geotherm (\citealp{Katsura2017}).

We performed additional MD simulations of Di\textsubscript{64}An\textsubscript{36} melts containing 0, 0.68, and 1.35 wt\% H\textsubscript{2}O at 2073 K and 0.5, 3, and 6 GPa. H\textsubscript{2}O decreases \emph{η} by \textasciitilde0.03 Pa s per 1 wt\% H\textsubscript{2}O---a \textasciitilde25\% reduction relative to the dry Di\textsubscript{64}An\textsubscript{36} \emph{η} of 0.09--0.11 Pa s---with a largely \emph{P}-independent effect. This dependence is weaker than observed by \citet{Chen2025} for Di melt (\textasciitilde40\% per 1 wt\% H\textsubscript{2}O), which may reflect the difference in melt polymerization between Di (NBO/T = 2.0) and Di\textsubscript{64}An\textsubscript{36} (NBO/T ≈ 0.94). Permeability is parameterized as \(k_\varphi = d^2\varphi^n/C\), where \emph{d} is grain size and \emph{n} and \emph{C} characterize melt-network geometry (\citealp{McKenzie1989}). We adopt \emph{n} = 2.6 and \emph{C} = 58 from Stokes flow simulations based on synchrotron x-ray microtomography of three-dimensional melt networks in synthetic olivine--basalt aggregates (\citealp{Miller2014}). Mantle grain size is assumed to vary with \emph{P} and \emph{T} following \citet{Behn2009}. Although stress and strain rate variations in the mantle may modify \emph{d}, \emph{n}, \emph{C}, and hence \emph{k\textsubscript{φ}}, and may depart from laboratory-derived values, our analysis focuses on the role of melt viscosity. This simplification may affect absolute estimates of \emph{V}\textsubscript{D} but preserves relative differences between scenarios, which remain informative.

Figure~\ref{fig:2}f shows melt mobility, expressed as Δ\emph{ρ}/\emph{η} and independent of mantle geometry. Under both wet and dry conditions, Δ\emph{ρ}/\emph{η} increases with depth, peaks at \textasciitilde150 km, and declines slightly at greater depths. This trend arises because \emph{η} decreases exponentially with depth---driven by rising \emph{T}---outpacing the more gradual reduction in buoyancy Δ\emph{ρ} as melt and solid densities converge. Hydration enhances melt mobility by nearly an order of magnitude. The melt segregation velocity, \emph{V}\textsubscript{D} (Figure~\ref{fig:2}g), exhibits a nearly identical depth dependence, reflecting the dominant control of \emph{η} on melt transport efficiency. Our calculations separate the intrinsic \emph{P} dependence of \emph{η} along an isotherm from its variation along a geotherm. Increasing \emph{P} raises \emph{η} past the isothermal minimum, whereas increasing \emph{T} lowers it; along the geotherm, \emph{η} falls with depth until the \emph{P} effect overtakes the \emph{T} effect. The resulting mobility maximum therefore reflects \emph{P}--\emph{T} competition along the geotherm, not the isothermal minimum near 3 GPa, and does not by itself establish ponding.

These results imply two key consequences: (1) efficient melt extraction at depth. High Δ\emph{ρ}/\emph{η} and \emph{V}\textsubscript{D} below \textasciitilde150 km promote rapid melt segregation, consistent with global receiver-function observations that place the base of the asthenosphere at depths of 150 km (\citealp{Hua2023}; Figure~\ref{fig:2}h); (2) melt accumulation beneath the lithosphere. The decline in both Δ\emph{ρ}/\emph{η} and \emph{V}\textsubscript{D} between the LAB and \textasciitilde150 km impedes upward melt transport, causing partial melts to accumulate and stagnate just below the LAB. This prediction aligns with geophysical evidence for a melt-rich layer at 45--70 km depth, including a high-conductivity zone in magnetotelluric data (\citealp{Naif2013}) and sharp, negative shear-wave velocity contrasts in high-frequency SS precursors (\citealp{Schmerr2012}). The calculated segregation velocities are idealized buoyancy-driven estimates for a fixed porosity, grain-size profile, and permeability law. They should not be read as observed or predicted ascent rates. They neglect porosity evolution, compaction-driven pressure gradients, and reaction-driven channel formation, which can alter melt transport rates and distribution (\citealp{McKenzie1984}).

We revisit the experiments of \citet{Sakamaki2013}. Although their \emph{η} measurements may be affected by experimental artifacts---as recently suggested by \citet{Russell2025}---our updated results still support their central implication that melt can pond at the LAB. However, the observed increase in Δ\emph{ρ}/\emph{η} does not require an \emph{η} minimum as originally proposed; instead, it is primarily driven by the exponential decrease in \emph{η} with depth controlled by rising \emph{T} along the geotherm.

To conclude, while basaltic melts do exhibit an intrinsic \emph{P}-induced \emph{η} minimum, this feature is neither necessary nor sufficient to produce melt accumulation at the LAB, which can be explained by thermally controlled \emph{η} alone. We adopt a simplified Di\textsubscript{64}An\textsubscript{36} composition and H\textsubscript{2}O as the sole volatile, which limits chemical realism but is necessary given the computational cost of NNP development. Extending this framework to multi-volatile systems and higher \emph{P--T} conditions is a critical next step for quantifying deep mantle melt transport, including across the mantle transition zone (\citealp{Fei2025}) and near the core--mantle boundary (\citealp{Boukare2025}).

\section*{Acknowledgments}

This work was supported by the Deep Earth Probe and Mineral Resources Exploration-National Science and Technology Major Project of China (no. 2025ZD1005508), the National Key Research and Development Program of China (grant no. 2023YFF0804100), the National Natural Science Foundation of China (no. 42225202, 42572036), the Natural Science Foundation of Hubei Province (no. 2025AFB454).

\section*{Open research}

All data supporting this study, including training data, neural network interatomic potentials, and example molecular dynamics simulation input files, are available from the corresponding author on reasonable request. Electronic structure calculations used VASP (version 5.4.4; \citealp{Kresse1996}). Molecular dynamics simulations used LAMMPS (version 2Aug2023; \citealp{Thompson2022}). Machine learning interatomic potentials were developed with DeePMD-kit (v2.2.10; \citealp{Wang2018}), DP-GEN (v0.10.2; \citealp{Zhang2020}), and GPUMD/NEP (v5.0; \citeauthor{Fan2021}, \citeyear{Fan2021}, \citeyear{Fan2022}). Thermodynamic equilibrium phase relations and thermoelastic properties of melt and minerals were computed using MAGEMin (version 1.8.5; \citealp{Riel2022}).

\section*{Conflict of interest disclosure}

The authors declare there are no conflicts of interest for this manuscript.


\clearpage
\noindent\begin{minipage}{\linewidth}
\begin{center}
\includegraphics[width=\linewidth,height=0.55\textheight,keepaspectratio]{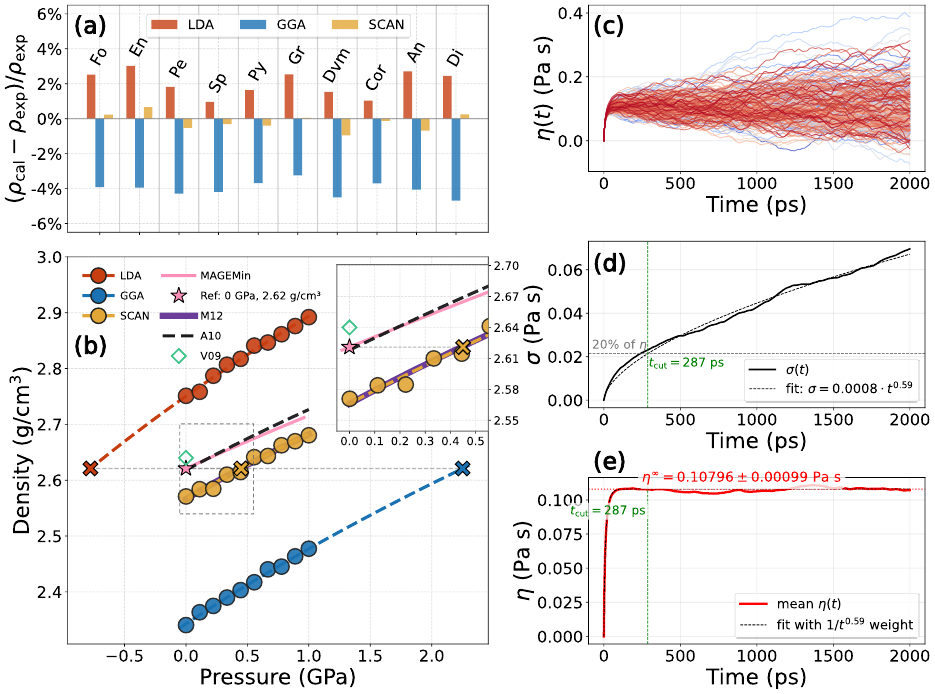}
\end{center}
\captionof{figure}{\textbf{Benchmarking mineral and melt density and illustrative Green--Kubo shear viscosity.} Left panel: \textbf{(a)} Density deviations of LDA-, GGA-, and SCAN-based neural network potentials (NNPs), trained on DFT reference data generated with the corresponding exchange--correlation functional, relative to experimental densities of ten representative crystalline silicate and oxide minerals at 0 GPa and 300 K. \textbf{(b)} Equation of state (EOS) of Di\textsubscript{64}An\textsubscript{36} melt at 1673 K. Densities calculated from LDA-, GGA-, and SCAN-based NNPs (filled circles) are fit to a second-order Birch--Murnaghan EOS (dashed lines) and extrapolated to the reference density (pink star: \emph{ρ} = 2.62 g cm\textsuperscript{-3} at 0 GPa, from oxide partial molar volumes; \citealp{Lange1997}); cross markers indicate these extrapolated intercepts. Also shown are the MAGEMin thermodynamic reference (\citealp{Riel2022}), the empirical MD EOS of \citet{Martin2012} (M12), the shock-compression EOS of \citet{Asimow2010} (A10), whose 0 GPa density is fixed to the \citet{Lange1997} reference value, and the density of \citet{Vuilleumier2009} (V09), adjusted to 1673 K. The inset details the 0--0.5 GPa region. Right panel: Shear viscosity of Di\textsubscript{64}An\textsubscript{36} melt computed via the Green--Kubo relation at 2073 K and 0 GPa. \textbf{(c)} Individual \emph{η}(\emph{t}) trajectories. The long-time stress autocorrelation function contains fewer independent samples, causing statistical oscillations in \emph{η}(\emph{t}). The final \emph{η} values were therefore obtained from the converged plateau region. \textbf{(d)} Convergence behavior revealed by the standard deviation \emph{σ}(\emph{t}). \textbf{(e)} Running mean \emph{η}(\emph{t}) and the long-time limit \emph{η}\textsuperscript{∞}, estimated using the time decomposition method (\citealp{Zhang2015}), yielding \emph{η}\textsuperscript{∞} = 0.1080 ± 0.0010 Pa s (bootstrap standard error).}\label{fig:1}
\end{minipage}\par
\clearpage
\noindent\begin{minipage}{\linewidth}
\begin{center}
\includegraphics[width=\linewidth,height=0.66\textheight,keepaspectratio]{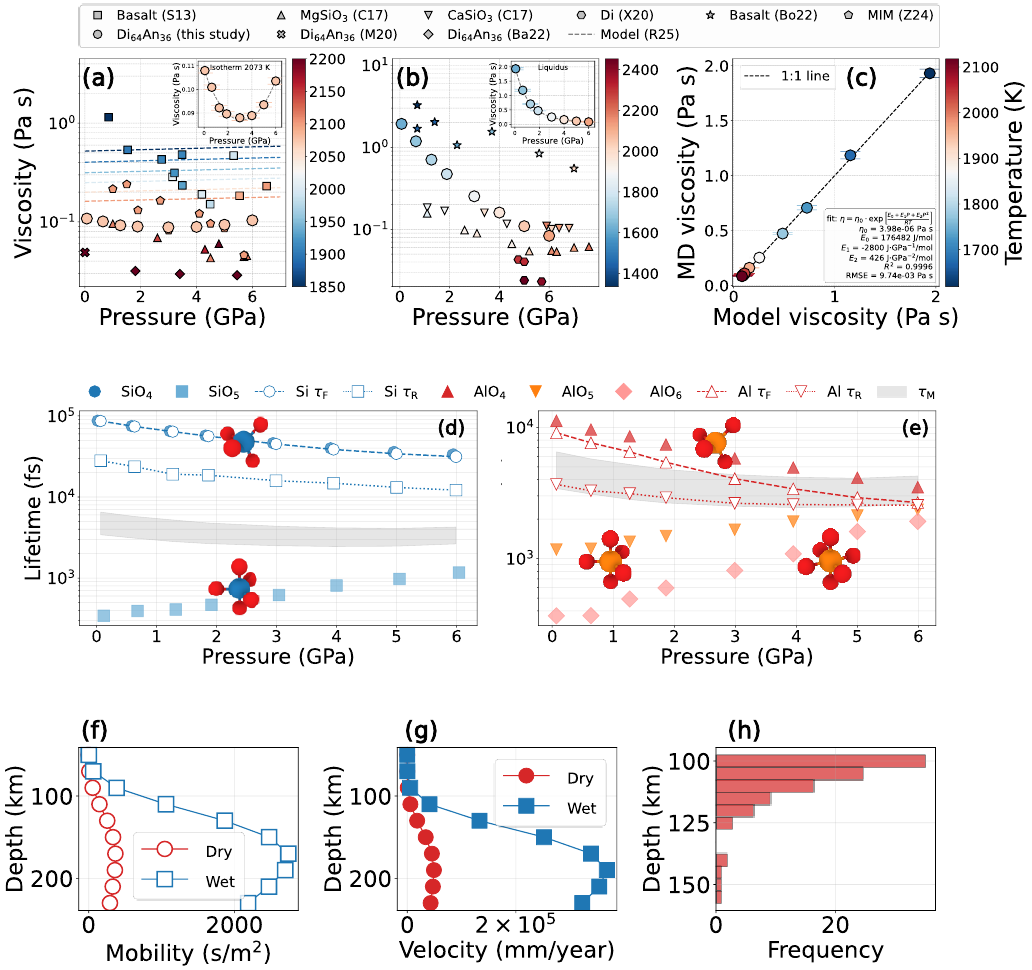}
\end{center}
\captionof{figure}{\textbf{Basalt viscosity, melt structure, and mobility in the asthenosphere. (a)} Nominal isothermal comparison near \textasciitilde2000 K; note that strictly isothermal conditions are experimentally challenging (see main text for further details). \textbf{(b)} Comparison along the liquidus. The inset in (a) highlights the U-shaped pressure dependence predicted at 2073 K, whereas the inset in (b) illustrates the apparent stronger pressure dependence arising from along-liquidus temperature variations. Previous results include Ba22 (\citealp{Bajgain2022}), Bo22 (\citealp{Bonechi2022}), C17 (\citealp{Cochain2017}), M20 (\citealp{Majumdar2020}), R25 (\citealp{Russell2025}), S13 (\citealp{Sakamaki2013}), X20 (\citealp{Xie2020}), and Z24 (\citealp{Zhou2024}). \textbf{(c)} Arrhenius model fit (Eq.~\ref{eq:2}) of all molecular dynamics (MD) viscosities across temperature and pressure. \textbf{(d,e)} Pressure dependence of Si--O (d) and Al--O (e) coordination lifetimes at 2073 K (Eq.~\ref{eq:3}). Solid symbols denote individual coordinations (SiO\textsubscript{4}, SiO\textsubscript{5}, AlO\textsubscript{4}, etc.); open symbols represent fraction-weighted (\emph{τ}\textsubscript{F}) and rate-weighted (\emph{τ}\textsubscript{R}) mean lifetimes. Grey bands denote bulk relaxation times estimated from the Maxwell model, \(\tau_{\mathrm M} = \eta/G^\infty\), with instantaneous shear modulus \emph{G}\textsuperscript{∞} = 20--40 GPa (\citealp{Dingwell1989}). \textbf{(f,g)} Melt mobility (Δ\emph{ρ}/\emph{η}; f) and segregation velocity (Eq.~\ref{eq:4}; g) as functions of depth, for dry (red) and hydrous (blue) conditions. \textbf{(h)} Histogram of global equilibration depths for primary magmas, compiled by \citet{Hua2023}.}\label{fig:2}
\end{minipage}\par

\clearpage

\end{document}